\documentclass[conference,transmag]{IEEEtran}
\usepackage{color}
\usepackage{ifpdf}
\usepackage{amsmath}
\usepackage{amssymb}
\usepackage{cite}
\usepackage{siunitx}

\ifCLASSINFOpdf
   \usepackage[pdftex]{graphicx}

\else

\fi

\newcommand{\mum}{\ensuremath{\mu\text{m}}}
\newcommand{\CMFS}{\ensuremath{\text{Co}_2\text{Mn}_{0.6}\text{Fe}_{0.4}\text{Si}}}
\newcommand{\hext}{\ensuremath{{\mu_0}{{H}_\text{ext}}}}
\newcommand{\Ith}{\ensuremath{j_\text{Th}}}
\newcommand{\uBLS}{\ensuremath{\mu}\text{BLS}}

\begin{document}

\title{Realization of a spin wave switch based on the Spin-Transfer-Torque effect}

\author{\IEEEauthorblockN{Thomas Meyer\IEEEauthorrefmark{1},
Thomas Br\"acher\IEEEauthorrefmark{1},
Frank Heussner\IEEEauthorrefmark{1}, 
Alexander A. Serga\IEEEauthorrefmark{1},
Hiroshi Naganuma\IEEEauthorrefmark{2},\\
Koki Mukaiyama\IEEEauthorrefmark{2},
Mikihiko Oogane\IEEEauthorrefmark{2},
Yasuo Ando\IEEEauthorrefmark{2},
Burkard Hillebrands\IEEEauthorrefmark{1,*}, and
Philipp Pirro\IEEEauthorrefmark{1}}

\IEEEauthorblockA{\IEEEauthorrefmark{1}Fachbereich Physik and Landesforschungszentrum OPTIMAS, Technische Universit\"at Kaiserslautern, \\67663 Kaiserslautern, Germany}
\IEEEauthorblockA{\IEEEauthorrefmark{2}Department of Applied Physics, Graduate School of Engineering, Tohoku University, Sendai 980-8579, Japan}
\IEEEauthorblockA{\IEEEauthorrefmark{*}IEEE Fellow}
\thanks{Corresponding author: T. Br\"acher (email: braecher@rhrk.uni-kl.de).}}

\markboth{IEEE Magnetics Letters}%
{Shell \MakeLowercase{\textit{et al.}}: Bare Demo of IEEEtran.cls for IEEE Transactions on Magnetics Journals}

\IEEEtitleabstractindextext{%

\begin{abstract} 
We investigate the amplification of externally excited spin waves via the Spin-Transfer-Torque~(STT)~effect in combination with the Spin-Hall-Effect~(SHE) employing short current pulses. The results reveal that, in the case of an overcompensation of the spin wave damping, a strong nonlinear shift of the spin wave frequency spectrum occurs. In particular, this shift affects the spin wave amplification using the SHE-STT effect. In contrast, this effect allows for the realization of a spin wave switch. By determining the {corresponding} working point, an efficient spin wave excitation is only possible in the presence of the SHE-STT effect yielding an increased spin wave intensity of a factor of 20 compared to the absence of the SHE-STT effect.
\end{abstract} 

\begin{IEEEkeywords}
Magnonics, Spin Electronics, Spin Torque.
\end{IEEEkeywords}}

\maketitle

\IEEEdisplaynontitleabstractindextext

\IEEEpeerreviewmaketitle
\section{Introduction}

\IEEEPARstart{O}{ver} the last years, in the research field of magnon spintronics, several concepts for information transport and logic devices were presented which are based on magnons, the quanta of spin waves, as information carriers~\cite{Khitun2008,Chumak2014,Vogt2014,Chumak2015,Davies2015,Wagner2016,Otani2018,Fischer2017,Heussner2017,Balinski2018}. Since magnons feature frequencies up to the THz regime while their wavelength is in the \mum\ and nm regime, they are a promising successor as information carrier to conventional electron-based CMOS technology. 

However, any magnon propagation is subject to damping, e.g., due to Gilbert damping~\cite{Gilbert2004}, which limits the magnon lifetime and, hence, the spin wave propagation length. Thus, in any future application based on extended magnonic circuits, an efficient amplification of spin waves is required. One potential approach for spin wave amplification is parametric amplification~\cite{Brächer2017} which allows for an efficient amplification of distinct spin wave modes. However, since magnonic devices enable wave-based computing, allowing for, e.g, a parallel operation employing spin waves at multiple frequencies, an amplification via parallel pumping requires multiple RF sources. 

In contrast, by employing the Spin-Transfer-Torque~(STT)~effect~\cite{Slonczewski1996} of a spin current injected into a magnetic layer, the effective spin wave damping can be manipulated for all magnon modes independent on their frequency. In combination with the Spin-Hall-Effect~(SHE)~\cite{Hirsch1999} to convert a charge current into a spin current in a normal metal, this allows for the manipulation of the effective spin wave damping which enables the control of the spin wave propagation length of externally excited spin waves~\cite{Gladii2016,Evelt2016}. In particular, in parallel computing devices employing multi-frequency operation, the SHE-STT effect allows for an easy control of the spin wave propagation.

In order to achieve effective spin wave amplification, it is important to note that the damping needs to be overcompensated. However, in this case, not only the information carrying spin waves, but also the thermal magnon density increases exponentially in time~\cite{Meyer2017-APL} which eventually drives the spin wave system into auto-oscillations~\cite{Hamadeh2014,Demidov2012,Demidov2014,Lauer2017,Kajiwara2010,Collet2016}. If the threshold for nonlinear magnon-magnon scattering processes is reached, this again limits the effective magnon lifetime and, hence, the spin wave propagation length. Thus, only in the case of short applied current pulses it might be possible to limit the total magnon density below the threshold for the onset of nonlinear magnon-magnon scattering events which might allow for an effective amplification of externally excited spin waves via the SHE-STT effect. Up to now, most of the reported experimental investigations on the SHE-STT effect have been performed using static applied charge currents or comparably long current pulses with pulse durations in the $\mu$s regime~{\cite{Hamadeh2014,Demidov2012,Demidov2014,Lauer2017,Kajiwara2010,Collet2016}}. Thus, no effective amplification of externally excited spin waves could be achieved.

In this Letter, we demonstrate that {the amplification of externally excited, propagating spin waves can be achieved} by using short current pulses. However, this amplification is limited due to a strong nonlinear frequency shift of the spin wave spectrum~\cite{Krivosik2010} which occurs due to the enhanced (thermal) magnon density. This nonlinear shift of the spin wave spectrum {changes the wave vector of the excited spin wave mode. This affects the spin wave excitation efficiency and, this way, limits the achievable spin wave amplification.}

Additionally, we show that this effect can be applied to realize a spin wave switch. For this, the excitation frequency and the external magnetic field are chosen such that, in the absence of a current pulse, the spin wave dispersion is above the excitation frequency. In this situation, no spin wave excitation is possible. However, if the spin wave damping is overcompensated and the magnon density increases, the {resulting} nonlinear frequency shift causes a lowering of the spin wave frequency spectrum. For an adequate selection of the excitation frequency and the externally applied magnetic field, spin waves can be efficiently excited if the excitation source and the charge current are applied simultaneously.

 

\section{Investigated sample and experimental setup}
\label{Sec:setup}
\begin{figure}[!t]
	 \centering
	 \includegraphics[width=0.4\textwidth]{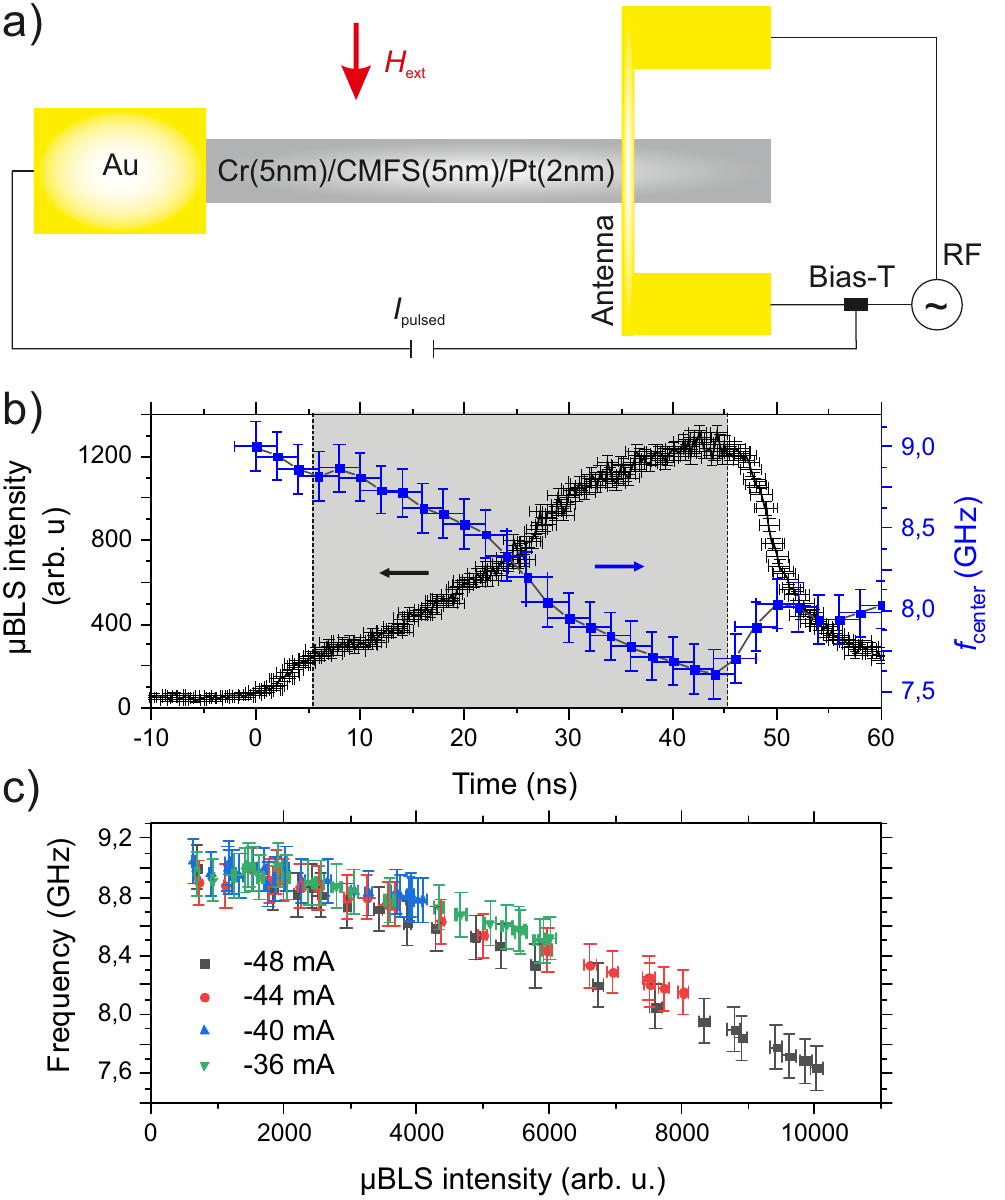}
	\caption{~(color online)~a)~Sketch of the investigated waveguide consisting of a Cr~($5\,\mathrm{nm}$)$\vert$\CMFS~($5\,\mathrm{nm}$)$\vert$Pt~($2\,\mathrm{nm}$) trilayer on a MgO substrate. {The right Au contact is connected to an RF source and acts as a spin wave antenna.} b)~Temporal evolution of the \uBLS\ intensity and the {corresponding} center frequency $f_\text{Center}$ of the spin wave spectrum for a \SI{50}{ns} long current pulse of \SI{-48}{mA}. The shading marks the area where the pulse exhibits a constant voltage amplitude, which is taken into account for the the evaluation applied in c. c)~spin wave intensity-dependent shift of $f_\text{Center}$ for different applied currents.}
	\label{Fig1}
\end{figure}

The investigated sample is sketched in Fig.~\ref{Fig1}a and consists of a Cr~(\SI{5}{nm})$\vert$\CMFS~(\SI{5}{nm})$\vert$Pt~(\SI{2}{nm}) trilayer which is grown on a MgO substrate by sputter deposition. The low-damping ferromagnetic full Heusler compound \CMFS~(CMFS)~\cite{Sebastian2012,Sebastian2015} constitutes the magnetic layer of the stack while the Cr layer and the Pt layer allow for the conversion of a charge current into a spin current via the SHE. Due to the opposite signs of the {respective} spin-Hall angles of Pt and Cr~\cite{Schreier2014,Du2014} in combination with the fact that the emerging spin currents enter the CMFS layer from opposite surfaces, the injected spin current is maximized while parasitic influences like, e.g., the Oersted fields of the charge currents partially compensate each other inside the CMFS layer. Thus, this trilayer design represents an efficient layer setup for the presented investigations on the SHE-STT effect~\cite{Meyer2017,Meyer2017-APL}.
%
%
%
%

Subsequently, the trilayer is patterned into \SI{7}{\mum} wide and \SI{30}{\mum} long rectangular slabs which act as spin wave waveguides. In order to apply charge currents, {a \SI{300}{nm} thick Au contact is fabricated at one end of the waveguide. The other end is contacted by a \SI{2}{\mum} wide microstrip antenna}. {In the experiment, an external magnetic field is applied along the antenna to magnetize the waveguide along its short axis.} By applying a radio frequency current (RF current) to the antenna, the {microwave magnetic fields originating from the RF current} excite spin waves at the {same} frequency at the position of the antenna. Subsequently, the spin waves propagate along the spin wave waveguide away from the antenna.

For the experimental investigation of the spin wave properties in the spin wave waveguide under the influence of the SHE-STT effect, time-resolved Brillouin light scattering microscopy~(\uBLS) is employed~\cite{Sebastian2015-BLS}. This technique allows for the determination of the density of incoherent thermal magnons which is the basis for the results shown in Sec.~\ref{Sec:f-shift}.
\section{Nonlinear frequency shift}
\label{Sec:f-shift}

Figure~\ref{Fig1}b exemplarily shows the time evolution of the \uBLS\ intensity~$I_{\mu \text{BLS}}$ (black curve) and the center frequency~$f_\text{Center}$ of the thermal spin wave spectrum (blue squares) in the center of the waveguide for a \SI{50}{ns} long current pulse of $j=-48~\text{mA}$ applied at $t = 0$ { and for an applied external magnetic field of $\mu_0H_\mathrm{ext}=70\,\mathrm{mT}$}. Here, the center frequency of the spectrum is determined as the weighted average~\cite{Meyer2017}. {The error bar in frequency corresponds to the channel width of $150\,\mathrm{MHz}$ used during the BLS measurement. The error bar in time of $1\,\mathrm{ns}$ corresponds to the actual time resolution of the BLS setup. The error bar in the intensity $I$ is given by $\sqrt{I}$.} For the investigated system, the threshold current, i.e., the current at which the spin wave damping is compensated by the SHE-STT effect, is \Ith~$ \approx -17~\text{mA}$. {This value of the threshold current was determined from the inverse rise time of the BLS intensity above the threshold, as is reported in our previous work Ref.\cite{Meyer2017-APL}}. For larger magnitudes of the applied current, the spin wave damping is overcompensated leading to a negative effective spin wave damping{, resulting in an} exponential increase of the magnon density during the current pulse~\cite{Meyer2017-APL}.

Considering the time-resolution of the experimental setup on the order of $1\,\mathrm{ns}$ and the rise- and fall times of the pulse of around \SI{5}{ns}, in the following, only the shaded region between the dashed lines in Fig.~\ref{Fig1}b is taken into account. Here, the increase of the spin wave intensity is accompanied by a monotonous decrease of $f_\text{Center}$. Both effects show a similar time scale until the end of the current pulse. To determine the origin of this frequency shift, during the current pulse, at each point in time, $f_\text{Center}$ is correlated {with} the {corresponding} spin wave intensity. The resulting intensity-dependent $f_\text{Center}$ for $j=-48~\text{mA}$ is depicted by the black squares in Fig.~\ref{Fig1}c. In addition, Fig.~\ref{Fig1}c shows the intensity-dependent center frequencies for different applied charge currents which reveal an almost identical behavior that is independent on $j$. This shows that the influence of Joule heating on the spin wave frequency spectrum is negligible. Furthermore, since $f_\text{Center}$ mainly depends on the magnon density, this indicates that a nonlinear frequency shift is the main origin of the observed frequency shift. Similar to~\cite{Krivosik2010}, the nonlinear frequency can be described by:
\begin{equation}
\label{Eq:nonlinearshift}
	f_\text{Center}=f_\text{Center,0}+\tilde{W} \vert c(t)\vert^2,
\end{equation}
with $f_\text{Center,0}$ indicating the center frequency in the absence of any nonlinear interactions, $\tilde{W}$ being the effective coefficient of the Hamiltonian describing the four-magnon processes and $c(t)$ denoting the effective spin wave amplitude, hence, $\vert c(t)\vert^2$ corresponds to the effective spin wave intensity.

\begin{figure}[!t]
	 \centering
	 \includegraphics[width=0.4\textwidth]{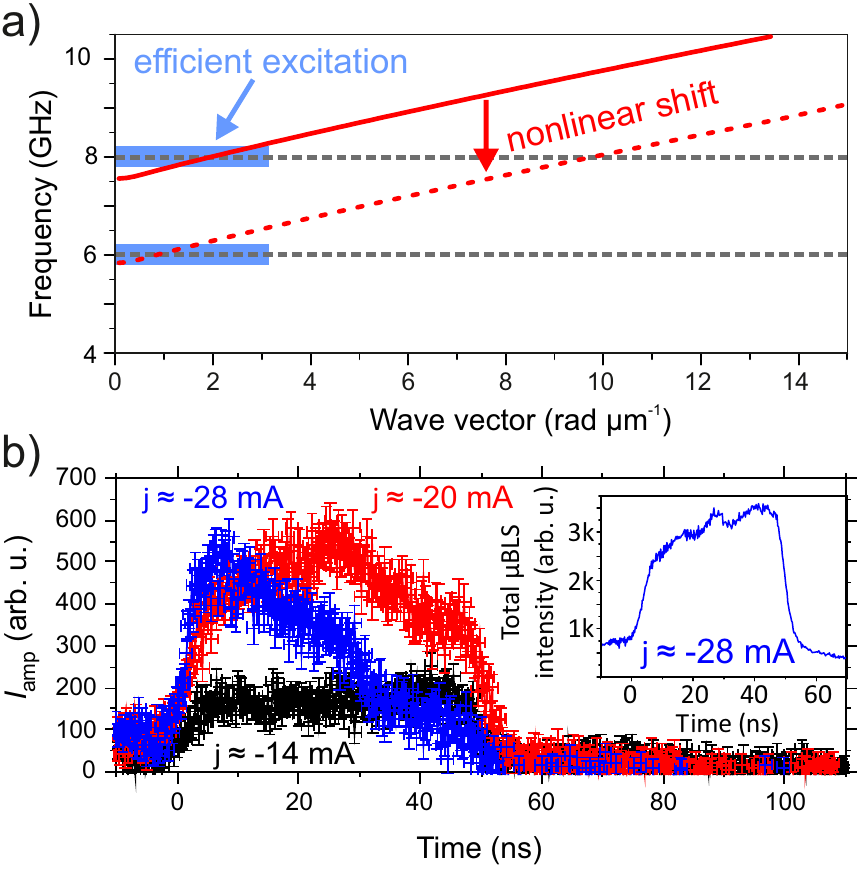}
	\caption{~(color online)~a)~spin wave dispersion for $\hext=52~\text{mT}$ (solid line). In the absence of a nonlinear shift, spin waves can efficiently be excited at \SI{8}{GHz} while no excitation is possible at a frequency of \SI{6}{GHz} (frequencies indicated by the horizontal dashed lines). The regions for an efficient excitation at \SI{6}{GHz} and \SI{8}{GHz} are indicated by the blue shaded areas, respectively. In contrast, if the dispersion is shifted due to a nonlinear shift (dashed red line), the wave vector of the excited spin wave increases causing a reduction of the excitation efficiency at \SI{8}{GHz} until no excitation is possible any more. In return, spin waves can be excited at \SI{6}{GHz} which can be employed to realize a spin wave switch.~b)~Temporal evolution of the influence of the SHE-STT effect on the intensity of externally excited spin waves at \SI{8}{GHz} for an applied current of \SI{-14}{mA} (black curve), \SI{-20}{mA} (red curve) and \SI{-28}{mA} (blue curve). The inset shows the total spin wave intensity integrated over the whole accessible spin wave frequency range for $j=-28~\text{mA}$.}
	\label{Fig2}
\end{figure}

As can easily be seen from Eq.~\ref{Eq:nonlinearshift}, with increasing spin wave intensity during the current pulse, the nonlinear shift of the spin wave dispersion as depicted in Fig.~\ref{Fig2}a causes a continuous increase of the excited spin wave wavevector during the current pulse. Considering a spin wave excitation via an antenna as discussed in Sec.~\ref{Sec:setup}, also the wavevector-dependent excitation efficiency needs to be taken into account. This efficiency is given by the Fourier-transform of the spatial extent of the antenna, i.e. its width, which corresponds to a maximal efficiency for small spin wave wavevectors as indicated by the blue shaded areas in Fig.~\ref{Fig2}a for an excitation frequency of \SI{6}{GHz} and \SI{8}{GHz}, respectively~\cite{Brächer2017-2}. Thus, the increase of the spin wave wave vector is the main limitation for the achievable amplification via the SHE-STT effect in the investigated system.
%
%
%
%

To show this in more detail, the influence of the SHE-STT effect on the intensity of externally excited spin waves at a frequency of \SI{8}{GHz} with an RF power of \SI{1}{mW} and for an externally applied magnetic field of $\hext=52~\text{mT}$ is investigated. The {corresponding} dispersion relation is shown in Fig.~\ref{Fig2}a. {At this field, the antenna can excite propagating spin waves efficiently in the absence of a nonlinear shift. To visualize the effect of the frequency shift} the time-resolved spin wave intensity at the excitation frequency \mbox{$I_\text{SHE-STT + excitation} (f=8~\text{GHz})$} at a distance of approximately $1~\mum$ to the antenna is determined for a simultaneous spin wave excitation and an applied current pulse. In addition, the spin wave intensity \mbox{$I_\text{SHE-STT only} (f=8~\text{GHz})$} caused by the SHE-STT effect in the absence of an external spin wave excitation and the spin wave intensity \mbox{$I_\text{excitation only} (f=8~\text{GHz})$} in the presence of external excitation without the application of a current pulse is determined. From this data, the influence of the SHE-STT effect on the excited spin waves can be derived via:
\begin{equation}
\label{Eq:interaction}
	I_\text{amp} = I_\text{SHE-STT + excitation} - (I_\text{SHE-STT only} + I_\text{excitation only}).
\end{equation}
The resulting temporal evolutions of $I_\text{amp}$ for different applied currents are depicted in Fig.~\ref{Fig2}b. For $j=-14~\text{mA}$ (black curve), an increase of the spin wave intensity which is constant in time is observed. On the one hand, this increase is caused by the increased spin wave propagation length causing an increased spin wave intensity at the position of the detection point. On the other hand, it should be noted that also the spin wave excitation efficiency at the antenna is enhanced if the effective spin wave damping is reduced yielding another contribution to the observed increase of the spin wave intensity. 

In contrast, if the spin wave damping is overcompensated, i.e., in the case of an exponential increase of the spin wave intensity during the current pulse, the nonlinear frequency shift limits the spin wave amplification. This is exemplarily shown by the red and the blue curves in Fig.~\ref{Fig2}b for $j=-20~\text{mA}$ and $j=-28~\text{mA}$, respectively. In both cases, at the onset of the current pulse, the spin wave intensity is significantly increased. Furthermore, for $j=-20~\text{mA}$, in the first half of the current pulse, an effective amplification of the externally excited spin waves can be achieved. In contrast, during the second half of the pulse, the nonlinear shift and the {consequent} lowering of the excitation efficiency caused by the exponentially increasing thermal magnon density  limit the amplification. This results in a monotonous decrease towards the end of the current pulse. For $j=-28~\text{mA}$, i.e., for a further decrease of the effective spin wave damping, already shortly after the onset of the current pulse, the nonlinear frequency shift causes a decrease of the intensity of the excited spin waves until it even vanishes at the end of the current pulse. This key result clearly shows the aforementioned limitation of an amplification of externally excited spin waves via the SHE-STT effect.

At this point, it is important to note that the decrease in $I_\text{amp}$ is caused by the aforementioned shift of the spin wave frequency spectrum and the resulting decrease in the excitation- and detection efficiency as indicated in Fig.~\ref{Fig2}a instead of a nonlinear saturation of the spin wave intensity. To show this, the inset in Fig.~\ref{Fig2}b depicts the spin wave intensity for $j=-28~\text{mA}$ integrated over the whole accessible spin wave frequency range. As can clearly be seen, the magnon density constantly increases in time. However, the dominant contribution to this intensity is the increased thermal magnon density due to the SHE-STT effect rather than the intensity of the externally excited spin waves.
\section{Realization of a spin wave switch}
\label{Sec:Switch}
In the previously presented investigations, the frequency of the externally excited spin waves is fixed to $f_\text{RF}=8~\text{GHz}$ which lies within the spin wave dispersion relation {in the absence of a nonlinear shift}. This allows, e.g., for an increase of the spin wave propagation length if the effective spin wave damping is reduced via the SHE-STT effect. However, for an effective spin wave amplification via an overcompensation of the spin wave damping, the occurrence of a nonlinear shift of the spin wave dispersion needs to be taken into account.

%
%
%
%
\begin{figure}[!t]
	 \centering
	 \includegraphics[width=0.4\textwidth]{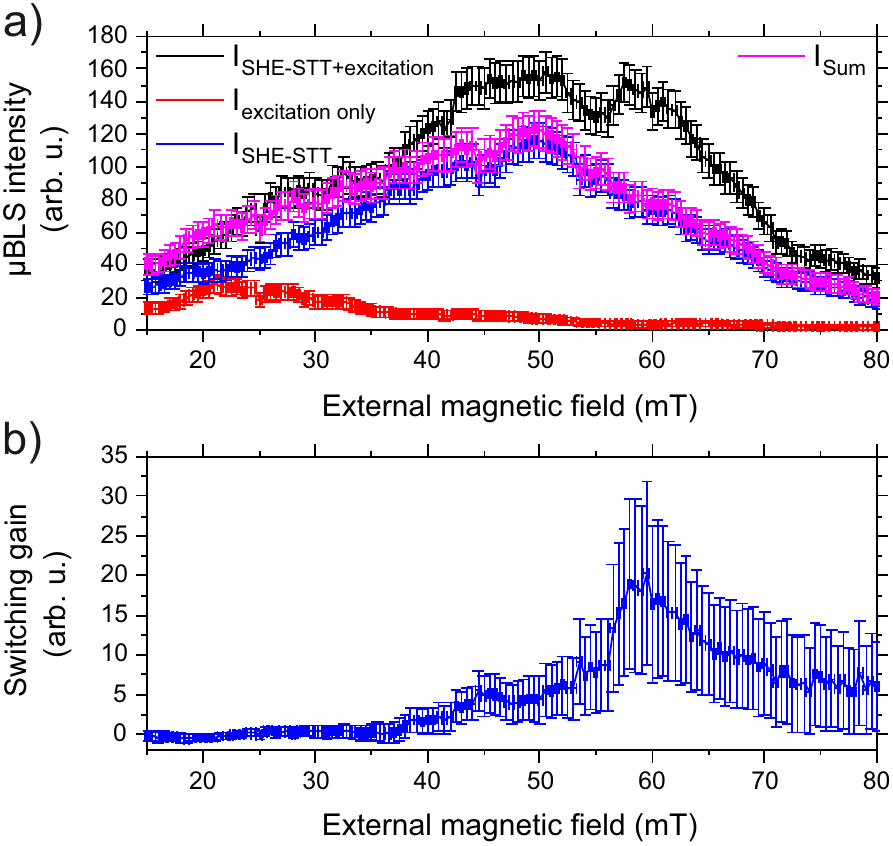}
	\caption{~(color online)~a)~Spin wave intensity at a frequency of \SI{6}{GHz} depending on the applied magnetic field in the case of the SHE-STT effect and the excitation being applied simultaneously (black curve), only in the presence of the SHE-STT effect (blue curve) and for the excitation in the absence of the SHE-STT effect (red curve). The magenta line indicates the sum of the red and the blue curve. b)~Switching gain of the externally excited spin waves in the presence of the SHE-STT effect according to Eq.~\ref{Eq:amplification}.}
	\label{Fig3}
\end{figure}
In return, in the following, this frequency shift is employed to realize a spin wave switch which only allows for an efficient spin wave excitation in the presence of a current pulse. For this, the excitation frequency is shifted below the spin wave dispersion relation. Thus, in the absence of a current pulse, no spin waves can be excited. If the effective spin wave damping is overcompensated by the SHE-STT effect, the spin wave dispersion is shifted into the frequency range of the excitation which enables an efficient spin wave excitation (see Fig.~\ref{Fig2}a). This should enable a very fast switching time since the relaxation of the spin wave dispersion should occur within the intrinsic magnon lifetime which is in the order of \SI{0.2}{ns} in the investigated system. {To demonstrate this concept, in the following we will apply a lower frequency of $6\,\mathrm{GHz}$. As indicated in Fig.~\ref{Fig2}a), the nonlinear shift can result in an efficient excitation of waves at this frequency, even at a high external field. This is the basis of the investigated spin wave switch.}

Assuming a spin wave detector in a potential application which is localized in a certain distance to the excitation source, the excited spin waves will only arrive at the detector if the excitation and the charge current pulse are present simultaneously. Due to the symmetry of the SHE-STT effect relative to the sign of the applied current, also the direction of the applied current plays an important role and represents an additional parameter to control the device. Only in the case of a decrease of the effective spin wave damping, spin waves can be excited and reach the detector.

To determine the operation point for the spin wave switch and to show that for a {proper} choice of the excitation frequency and the external field \hext, spin waves can only be efficiently excited during an applied current pulse, \uBLS\ measurements are performed for varying \hext\ while the excitation frequency is fixed at \SI{6}{GHz} with an RF power of \SI{2}{mW}. The experimentally obtained spin wave intensity \mbox{$I_\text{SHE-STT + excitation} (f=6~\text{GHz})$} for an applied current of approximately \SI{-45}{mA} is depicted by the black curve in Fig.~\ref{Fig3}a. In analogy to what we presented above. In order to determine the influence of the SHE-STT effect on the externally excited spin waves, the spin wave intensity solely caused by the SHE-STT effect in the absence of an external spin wave excitation \mbox{$I_\text{SHE-STT only} (f=6~\text{GHz})$} (blue curve) and the spin wave intensity given solely by the excitation source in the absence of the SHE-STT effect, \mbox{$I_\text{excitation only} (f=6~\text{GHz})$} (red curve), is depicted. The sum of the two single contributions \mbox{$I_\text{SHE-STT only} + I_\text{excitation only}$} is indicated by the magenta line. 

As expected, in the absence of the SHE-STT effect, the most efficient excitation of spin waves occurs at rather small magnetic fields around \hext$=25~\text{mT}$. In contrast, if the SHE-STT effect and the external spin wave excitation are present at the same time, the most efficient excitation is obtained at much larger magnetic fields due to the lowering of the spin wave dispersion during the current pulse. To determine the working point and the {switching gain,~$SG$,} of externally excited spin waves via the SHE-STT effect, the ratio
\begin{equation}
\label{Eq:amplification}
 SG=\frac{I_\text{SHE-STT + excitation} - (I_\text{SHE-STT only} + I_\text{excitation only})}{I_\text{excitation only}}
\end{equation}
is determined in dependence on \hext\ and shown in Fig.~\ref{Fig3}b.

The result reveals a maximum obtained switching {gain} of the externally excited spin waves of {$20\pm11$} at a magnetic field of approximately \SI{60}{mT}. This shows that the SHE-STT effect can be employed as a spin wave switch. Furthermore, due to the operation using charge current, this device allows for an easy implementation of a spin wave switch into conventional CMOS circuits.

\section{Conclusion}
The presented experimental investigations reveal a strong nonlinear frequency shift as a consequence of the enhanced magnon density due to a damping compensation by the SHE-STT effect. This limits a direct effective spin wave amplification via the SHE-STT effect. In contrast, in Sec.~\ref{Sec:Switch}, it could be shown that this effect can be employed to realize a spin wave switch which allows for an efficient excitation of spin waves only if the external excitation and the SHE-STT effect are simultaneously present in the system. Furthermore, since this switch is based on a nonlinear shift, the switching time should be rather fast since the relaxation of the spin wave frequency spectrum should occur within the intrinsic magnon lifetime. This is an important requisite for any future application using magnons and prefers a very short magnon lifetime. Since the presented approach is based on excited magnons and charge currents flowing through the waveguide and its adjacent layers, it can be implemented into any magnonic network and allows for an easy combination with CMOS technology.

\section*{Acknowledgment}
The authors gratefully acknowledge financial support by the DFG in the framework of the Research Unit TRR 173 ``Spin+X'' (Projects B01 and B04) and by the DFG Research Unit 1464 and the Strategic Japanese-German Joint Research Program from JST: ASPIMATT.

\ifCLASSOPTIONcaptionsoff
  \newpage
\fi


\end{document}